                                \documentclass[nohyper,12pt,letterpaper]{JHEP}
                                \usepackage[dvips]{epsfig}
                                \usepackage{epsf,amsfonts,amssymb}

                                \newcommand{\be}{\begin{equation}}
                                \newcommand{\ee}{\end{equation}}
                                \newcommand{\ben}{\begin{displaymath}}
                                \newcommand{\een}{\end{displaymath}}
                                \newcommand{\bea}{\begin{eqnarray}}
                                \newcommand{\eea}{\end{eqnarray}}
                                \newcommand{\bean}{\begin{eqnarray*}}
                                \newcommand{\eean}{\end{eqnarray*}}
                                \newcommand{\nn}{\nonumber \\}
                                \newcommand{\ba}{\begin{array}}
                                \newcommand{\ea}{\end{array}}
                                \newcommand{\bi}{\begin{itemize}}
                                \newcommand{\ei}{\end{itemize}}

                                \font\mybb=msbm10 at 10pt
                                \def\bb#1{\hbox{\mybb#1}}
                                
                                \def\bR {\bb{R}}
                                \def\bE {\bb{E}}

                                \newcommand{\sect}[1]{\setcounter{equation}{0}\section{#1}}

                                \title{The `s-rule' exclusion principle and vacuum
                interpolation in
                                worldvolume dynamics}

                                \author{Joaquim Gomis\\
                                Departament ECM, Facultat de F{\'\i}sica,\\
                                Institut de F{\'\i}sica d'Altes Energies and \\
                                CER for Astrophysics, Particle Physics and Cosmology,\\
                                Universitat de Barcelona, Diagonal 647,\\
                                E-08028 Barcelona, Spain}

                                \author{Paul K.\ Townsend and Mattias N.R. Wohlfarth\\
                                   Department of Applied Mathematics and Theoretical Physics\\
                                   Centre for Mathematical Sciences, Wilberforce Road, \\
                                   Cambridge CB3 0WA, U.K.}

                                \abstract{We show how the worldvolume realization of the
                Hanany-Witten
                                effect for a supersymmetric D5-brane in a D3 background also
                provides a classical
                                realization of the `s-rule' exclusion
                                principle. Despite the supersymmetry, the force on the
                                D5-brane vanishes only in the D5 `ground state', which
                                is shown to interpolate between 6-dimensional Minkowski space
                                and an $OSp(4^*|4)$-invariant $adS_2\times S^4$ geometry.
                                The M-theory analogue of these results is briefly discussed.}


                                \preprint{DAMTP-2002-131 \\ \tt{hep-th/0211020}}

\begin{document}


                                \sect{Introduction}

                                The worldvolume dynamics of a probe brane in the supergravity
                                background of another brane provides a useful way to
                understand
                                certain
                                interactions between the two string/M theory branes. Consider
                the case
                                of a
                                D5-brane in the presence of N coincident D3-branes, aligned so
                as to
                                preserve
                                1/4 supersymmetry according to the array
                                \bea
                                \label{array1}
                                \begin{array}{lccccccccc}
                                N\ D3: & 1 & 2 & 3 & - & - & - & - & - & - \\
                                probe\, D5: & - & - & - & 4 & 5 & 6 & 7 & 8 & -
                                \end{array}
                                \eea
                                For large N we may replace the D3-branes by the supergravity
                D3-brane
                                since the curvature of this 1/2 supersymmetric solution of IIB
                                supergravity is everywhere small in this limit, and the
                constant
                                dilaton
                                may be chosen such that the IIB string coupling is small. We
                should
                                therefore expect the worldvolume dynamics of a probe D5-brane
                in this
                                spacetime to capture effects associated to the interaction of
                the
                                D5-brane
                                with the N D3-branes.

                                One such effect was pointed out (in a dual context) by Hanany
                and
                                Witten
                                \cite{HW}: if the D5-brane is initially separated from the
                D3-branes
                                along
                                the
                                $9$-axis then a IIB string stretching between the D5-brane and
                each of
                                the N
                                D3-branes is created as the D5-brane is pulled through the
                D3-branes.
                                The
                                final, and still 1/4 supersymmetric, brane-plus-string
                configuration
                                can be
                                represented by the array
                                \bea
                                \label{array2}
                                \begin{array}{lccccccccc}
                                N\, D3: & 1 & 2 & 3 & - & - & - & - & - & - \\
                                probe\, D5: & - & - & - & 4 & 5 & 6 & 7 & 8 & - \\
                                N\ strings: & - & - & - & - & - & - & - & - & 9
                                \end{array}
                                \eea
                                The worldvolume description of this `string creation' effect
                in terms
                                of the
                                worldvolume dynamics of the D5-brane was initiated by Callan
                et al.
                                \cite{CGS}, who solved an equation found by Imamura for an
                                $S^5$-wrapped D5-brane in $adS_5\times S^5$ \cite{Imamura},
                which is
                                the
                                near-horizon limit of the D3-brane geometry \cite{GT}.  This
                provides
                                a
                                worldvolume realization of Witten's baryon vertex
                                \cite{baryonvertex} but, for reasons that we will explain
                later, the
                                Hanany-Witten (HW) effect can only be properly understood in
                the
                                context of
                                the {\sl full} D3 geometry. An extension of Imamura's equation
                to the
                                full
                                D3-brane geometry was also proposed and analysed numerically
                in
                                \cite{CGS},
                                but the status of this equation only became clear in
                subsequent works
                                in
                                which it was recovered from the conditions of minimal energy
                                \cite{CGMV} and
                                preservation of 1/4 supersymmetry \cite{GRST}, and solved
                analytically
                                \cite{CRS}. 

                                Note that the array (\ref{array2}) corresponds to a
                supersymmetry
                                preserving configuration only for one orientation of the N
                strings;
                                given
                                the orientations of the D3-branes and the probe D5-brane
                (i.e., a
                                choice
                                of brane vs. anti-brane in each case) only one of the two
                possible
                                string
                                orientations (string or anti-string) is compatible with
                supersymmetry.
                                As
                                the D5-brane is passed through the D3-branes, starting from
                the
                                configuration with N strings represented by (\ref{array2}),
                the
                                orientation of the strings that connect them would be
                reversed, and
                                hence
                                supersymmetry would be broken, if it were not for the fact
                that these
                                strings are destroyed by the `reverse' Hanany-Witten effect,
                which
                                returns us to the configuration without strings represented by
                the
                                array
                                (\ref{array1}). This escape from contradiction fails if any of
                the
                                D5-branes is connected to a D3-brane by more than one string,
                and this
                                led Hanany and Witten to propose (for their dual brane setups)
                that
                                any
                                such multi-string configuration would break supersymmetry
                \cite{HW}.
                                This `s-rule' can be understood as a quantum effect in IIB
                superstring
                                theory: the ground state of a string stretched between a
                D5-brane and
                                a
                                totally orthogonal D3-brane is fermionic, so the Pauli
                exclusion
                                principle
                                forces any additional strings into non-supersymmetric states
                of higher
                                energy \cite{BG}. One aim of this paper is to show how the
                s-rule
                                also
                                has a {\sl classical} explanation in terms of D5-brane
                worldvolume
                                dynamics; this is similar in spirit (but quite different in
                detail) to
                                the classical interpretation of the T-dual $D2\bot D6$ s-rule
                in terms
                                of the worldvolume dynamics of M2-branes in M-theory \cite{BG}
                (see
                                also \cite{MY}).

                                As just explained, the equations relevant to the $D5\bot D3$
                setup
                                under
                                consideration here have been found and solved in previous
                studies of
                                the
                                HW effect. However, these previous analyses are incomplete in
                several
                                respects. To explain why we must describe some features of the
                                function
                                $Z(\rho)$ that gives the position $Z$ on the 9-axis of the D5
                probe as
                                a
                                function of radial distance $\rho$ on the probe. This function
                depends
                                on
                                two
                                integration constants, a distance $Z_\infty$ which gives the
                                separation of
                                the asymptotic planar D5-brane from the D3-branes, and a
                constant
                                $\nu$ that is linearly related to the Born-Infeld
                                (BI) electric charge as measured by the flux at infinity.
                                Specifically, $Z(\rho)$ is given implicitly by the
                                equation\footnote{The $\arctan$ function here takes values in
                the
                                interval $[0,\pi]$.}
                                \be
                                \label{finald5}
                                Z= Z_\infty + {L^4 \over 2 \rho^3} \left[ \arctan
                \left({\rho\over
                                Z}\right)
                                -
                                {\rho Z \over \rho^2 + Z^2} - \pi \nu\right]
                                \ee
                                where $L$ is the `size' of the D3-brane core. For $0\le \nu
                \le
                                1$,
                                the function $Z(\rho)$ describes, for {\sl positive}
                $Z_\infty$, a
                                D5-brane
                                connected by $\nu N$ strings to the D3-branes; the $\nu=0$
                case
                                corresponds
                                to the array (\ref{array1}) (with no strings) and the $\nu=1$
                case to
                                the
                                array (\ref{array2}) (with $N$ strings). These two solutions
                are
                                interchanged
                                by taking $Z_\infty \rightarrow -Z_\infty$, so they actually
                belong to
                                a
                                single {\sl family} of solutions (depending on $Z_\infty$)
                that
                                provides a
                                worldvolume realization of the HW effect. Analogous families
                of
                                solutions
                                with $\nu>1$ have not been considered previously. This neglect
                was
                                possibly
                                motivated by the s-rule, which might seem to suggest that
                solutions
                                with
                                $\nu>1$ must be unphysical. However, we shall show here that
                the
                                $\nu>1$
                                solutions have a simple physical interpretation, again in
                terms of
                                $\nu N$
                                strings emerging from the D5-brane, but at most $N$ of these
                strings
                                end on
                                the D3-brane, thus  confirming the s-rule.

                                Although integer $\nu$ yields D5-brane geometries that have a
                simple
                                interpretation in terms of attached strings, non-integer $\nu$
                is also
                                possible because the D5-brane is infinite. Taking into account
                the
                                freedom
                                represented by the integration constant $Z_\infty$, one can
                restrict
                                $\nu$ to the range
                                \be
                                \nu \ge {1\over2}
                                \ee
                                without loss of generality. The minimal $\nu=1/2$ case is of
                                particular
                                interest but many of its special properties have been
                overlooked
                                previously. Another aim of this paper is to provide a more
                complete
                                treatment of this case. In particular,
                                we show that the induced metric on the D5-brane interpolates
                between
                                six-dimensional Minkowski space (as $\rho\rightarrow
                                \infty$) and $adS_2\times S^4$ (as $\rho\rightarrow 0$).
                Moreover, the
                                BI fields vanish in both limits, so we have a
                                vacuum interpolation `on the brane' analogous to the
                interpolation
                                noted in
                                \cite{GT} for the D3 background. In confirmation of this
                                interpretation we
                                show that the $adS_2\times S^4$ D5-brane vacuum has double the
                number
                                of
                                supersymmetries of the interpolating D5-brane. In fact, it is
                                invariant
                                under the transformations generated by the supergroup
                $OSp(4^*|4)$,
                                which
                                was identified in \cite{CGMV} as the `ground-state' supergroup
                                but without proper identification of the corresponding
                                D5-brane configuration.

                                We should note at this point that the $adS_2\times S^4$
                D5-vacuum has
                                recently been discussed \cite{ST,NMT} in the context of the
                adS/dCFT
                                correspondence \cite{KR,dWFO}; it can be viewed as the
                near-D3-horizon
                                limit
                                of the $Z_\infty=0$ case of a $\nu=1/2$ D5-brane; the
                $OSp(4^*|4)$
                                symmetry can be interpreted as conformal supersymmetry for an
                                ${\cal N}=8$ superconformal quantum mechanics. By taking
                                $Z_\infty\ne0$ we
                                then find supersymmetric (but non-conformal) deformations of
                                $adS_2\times
                                S^4$ with potential implications for the adS/dCFT
                                correspondence.

                                In addition to providing a much simplified derivation of
                                (\ref{finald5}) from supersymmetry, we also present a
                simplified
                                formula for the energy density, which we use to interpret the
                results
                                of our analysis of (\ref{finald5}), and a new formula for the
                net force
                                exerted on the D5-brane by the D3-branes. Surprisingly, this
                force 
                                does not generally vanish despite the fact that the D5-brane
                                configuration is both static and
                                supersymmetric! This is possible because a finite force cannot
                move an
                                object of infinite mass such as an infinite 5-brane. Whenever
                it would be 
                                possible to compactify the D5-brane (for example, by periodic
                                identification) then the force must vanish because the total
                mass on
                                which it acts would then be finite and
                                the D5-brane otherwise could not
                be
                                static. Such a compactification is possible only if $\nu=1/2$
                (because 
                                only in this case is there no BI flux at infinity). We
                                find that the force vanishes precisely in this case,
                                but not otherwise.

                                We will begin with a re-derivation of the result
                (\ref{finald5}) from
                                supersymmetry that incorporates significant simplifications,
                mainly
                                due to a
                                better gauge choice. We then reconsider the energy of
                                the D5-brane and compute the force on it. 
                                In the subsequent section we review the worldvolume
                                interpretation of the HW effect for the $\nu=1$ case and
                extend the
                                analysis
                                to $\nu>1$; this yields our worldvolume interpretation of the
                s-rule.
                                We
                                then turn to the $\nu=1/2$ case and demonstrate its vacuum
                                interpolation
                                property, and the enhanced supersymmetry of the $adS_2\times
                S^4$
                                embedding in $adS_5\times S^5$. We leave to a final section a
                                discussion
                                of how similar results apply to a probe M5-brane in an M5
                background,
                                and implications for adS/dCFT.

                                \section{Baryonic D5-brane revisited}
                                \label{d5revisit}

                                We consider a 1/2 supersymmetric D3 background solution of IIB
                                supergravity
                                for which the dilaton is constant and the only non-zero fields
                are the
                                metric and Ramond-Ramond (RR) 5-form field strength $R_5$.
                Choosing
                                cylindrical polar coordinates for the transverse $\bE^6$
                space, we
                                have the
                                metric
                                \be
                                ds^2_{10} = U^{-1/2} \left[-dT^2 + d\vec X \cdot d\vec X
                \right] +
                                U^{1/2}\left[ d\Upsilon^2 + \Upsilon^2 d\Omega_4^2(\Xi) + dZ^2
                \right]
                                \ee
                                where $\vec X$ are $\bE^3$ cartesian coordinates, $\Upsilon$
                is the
                                radial
                                coordinate in $\bE^5$ and $d\Omega_4^2(\Xi)$ is the
                $SO(5)$-invariant
                                metric on the unit 4-sphere, parametrized by four angles
                $\{\Xi\}$.
                                The
                                function $U$ is given by
                                \be
                                \label{ufun}
                                U= 1 + {L^4\over (\Upsilon^2 + Z^2)^2}\, 
                                \ee
                                where the D3-brane core size $L$ is given in terms of the
                integer $N$,
                                the IIB
                                string coupling constant $g_s$ and the `fundamental' IIB
                string
                                tension $T_f$ by
                                \be
                                \label{L}
                                L^4 = {g_sN \over \pi T_f^2}\, .
                                \ee
                                The RR 5-form is 
                                \be
                                R_5 = 4L^4\left[\omega_5 + \star \omega_5\right]
                                \ee
                                in polar coordinates for $\bE^6$, where $\omega_5$ is the
                volume
                                5-form on
                                the
                                unit  5-sphere and $\star \omega_5$ is its 10-dimensional
                Hodge dual.
                                In our cylindrical polar coordinates,
                                \bea
                                \label{om5}
                                \omega_5 &=& \sin^4\Theta\ d\Theta\wedge \omega_4 \nn
                                &=& {3\over8} d\left[\Theta -\sin\Theta\cos\Theta -{2\over3}
                                \sin^3\Theta\cos\Theta\right]\wedge \omega_4
                                \eea
                                where $\omega_4$ is the volume 4-form on the unit 4-sphere,
                and
                                \be
                                \tan\Theta = \Upsilon/Z\, .
                                \ee

                                Given an asymptotically flat D5-brane in this D3 background,
                we may
                                choose
                                worldvolume coordinates $x^i=(t,\rho,\xi)$, where $\{\xi\}$ are
                four
                                angular
                                coordinates for the 4-sphere at fixed radial distance $\rho$
                from a
                                worldspace origin. The worldvolume diffeomorphisms may now be
                fixed by
                                the
                                gauge choice
                                \be
                                T=t,\qquad \Upsilon =\rho,\qquad \{\Xi\} =\{\xi\} \, .
                                \ee
                                This leaves  $\vec X$ and $Z$ as the worldvolume fields
                determining
                                the
                                geometry of the D5-brane. Given the static $SO(5)$-invariant
                ansatz
                                \be
                                \vec X\equiv 0 , \qquad Z= Z(\rho)\, ,
                                \ee
                                the induced worldvolume metric $g$ is
                                \be
                                ds^2(g)= - U^{-1/2}(\rho)\ dt^2 + U^{1/2}(\rho)\left\{\left[1
                +
                                (Z')^2\right]
                                d\rho^2 + \rho^2 d\Omega_4^2(\xi)\right\}
                                \ee
                                where the prime indicates differentiation with respect to
                $\rho$ and,
                                now,
                                \be
                                \label{Unow}
                                U(\rho)= 1\ + \ {L^4\over \left[\rho^2 + Z^2(\rho)\right]^2}\,
                .
                                \ee
                                We must also take into account the worldvolume Born-Infeld
                field
                                strength
                                $F=dV$. Given the ansatz
                                \be
                                V= \Phi(\rho) dt
                                \ee
                                we have a radial electric field $E=\Phi'$.

                                Let ${\cal R}_5$  be the pullback of
                $R_5$ to the worldvolume;
                the
                                D5-brane worldvolume action in the chosen background is then
                                \be
                                \label{d5action}
                                I= -T_5\int d^6x \, \sqrt{-\det (g+ F)} \  + \ T_5\int {\cal
                                R}_5 \wedge V\, ,
                                \ee
                                where $T_5$ is the D5-brane tension, given in terms of the
                inverse
                                string
                                tension and the string coupling constant $g_s$ by
                                \be
                                \label{t5}
                                T_5 = {T_f^3\over 4\pi^2 g_s}\, .
                                \ee
                                For our gauge choice and ansatz we have
                                \be
                                \label{dbilag}
                                \sqrt{-\det (g+ F)} =  \rho^4 U\sqrt{1+ (Z')^2 -E^2}\
                vol_4(\xi)
                                \ee
                                where $vol_4(\xi)$ is the volume scalar density on the unit
                4-sphere.
                                To similarly simplify the remaining (Wess-Zumino) term in the
                action,
                                we first observe that the pullback of $\star \omega_5$
                vanishes
                                because
                                $d\vec X\equiv 0$; then, using (\ref{om5}), we find that
                                \be
                                {\cal R}_5 \wedge V = -4 L^4 \Phi\ dt\wedge d\left[\theta -
                                \sin\theta \cos\theta -{2\over3}\sin^3\theta
                \cos\theta\right]\wedge
                                \omega_4
                                \ee
                                where the function $\theta(\rho)$ is
                                determined in terms of $Z(\rho)$ through the
                                relation\footnote{A similar function was introduced in
                                \cite{CPR} for a D5-brane in $adS_5\times S^5$.}
                                \be
                                \label{ztheta}
                                \tan\theta = {\rho\over Z} \, .
                                \ee
                                The integral over the angular variables $\{\xi\}$ in
                (\ref{d5action}) is
                                now trivially done and yields a factor of $8\pi^2/3$ (this
                being the
                                volume of the unit 4-sphere); discarding a total derivative,
                                we are then left with the effective Lagrangian density
                                \be
                                \label{lagd5}
                                {\cal L} = - {8\pi^2T_5\over3}\left\{\rho^4 U \sqrt{1+ (Z')^2
                -E^2} -
                                {3\over 2} L^4 E \left[\theta - \sin\theta \cos\theta
                                -{2\over3}\sin^3\theta
                                \cos\theta\right] \right\} .
                                \ee
                                The equation of motion for the BI field $\Phi$ yields the
                Gauss law
                                constraint for $E$, which can be integrated immediately to
                give
                                \be
                                \label{gl}
                                {U\rho^4 E\over \sqrt{1+ (Z')^2 -E^2}}  =  {3\over2} L^4
                                \left[\pi\nu - \theta + \sin\theta \cos\theta
                                + {2\over3}\sin^3\theta\cos\theta\right]\, ,
                                \ee
                                where $\nu$ is an integration constant.

                                We now proceed to determine the conditions required by partial
                                preservation
                                of supersymmetry.  Let $\Gamma_A= (\Gamma_T,
                                \Gamma_{\vec X},
                                \Gamma_\Upsilon, \Gamma_Z, \Gamma_1, \Gamma_2, \Gamma_3,
                \Gamma_4)$ be
                                constant Dirac matrices, with the last four associated in the
                standard
                                way to the four angles $\{\Xi\}$ parametrizing $S^4$. Let $\chi$
                be a
                                covariantly constant $Sl(2;\bR)$-doublet chiral spinor in the
                D3
                                background.
                                Such spinors take the form
                                \be
                                \label{chiform}
                                \chi = U^{-1/8} \epsilon
                                \ee
                                where $\epsilon$ is a {\sl Minkowski} space covariantly
                constant
                                spinor
                                subject to the `$D3$ constraint'
                                \be
                                i\sigma_2 \otimes
                                \Gamma_T\Gamma_{X_1}\Gamma_{X_2}\Gamma_{X_3}\ \epsilon
                                =\epsilon\, ,
                                \ee 
                                where $\sigma_2$ is the $2\times 2$ Pauli matrix.
                                Owing to the chiral nature of $\epsilon$, this is equivalent
                to
                                \be
                                \label{d3con}
                                i\sigma_2\otimes \Gamma_Z \Gamma_\Upsilon \Gamma_*\ \epsilon
                =\epsilon\,
                                ,
                                \ee
                                where 
                                \be
                                \Gamma_* = \Gamma_1\Gamma_2\Gamma_3\Gamma_4\, .
                                \ee
                                Note that $\Gamma_*$ commutes with $\Gamma_T$,
                $\Gamma_\Upsilon$ and
                                $\Gamma_Z$, and is such that $\Gamma_*^2=1$.

                                In the presence of a D5-brane there is an additional
                constraint on
                                $\epsilon$ of the form $\Gamma_\kappa\epsilon =\epsilon$,
                where
                                $\Gamma_\kappa$ is the kappa-symmetry matrix of the super
                D5-brane.
                                In the conventions of \cite{BT}, and for a purely electric BI
                field,
                                this
                                additional condition is
                                \be\label{gamep}
                                \sqrt{-\det(g+ F)} \ \epsilon = \left[\sigma_1\otimes
                                \Sigma - iE \sigma_2 \otimes \gamma^t \gamma^\rho \Sigma
                                \right]\epsilon
                                \ee
                                where $\gamma_i$ are the induced
                                worldvolume Dirac matrices, and
                                \be
                                \Sigma = {1\over 6!} \varepsilon^{ijklmn}
                                \gamma_i\gamma_j\gamma_k\gamma_l\gamma_m\gamma_n\, .
                                \ee
                                Given spacetime frame 1-forms $E^A =dX^M E_M{}^A$, we have
                $\gamma_i =
                                \partial_iX^M E_M{}^A \Gamma_A$. For the obvious choice of
                zehn-bein
                                $E_M{}^A$ we find that
                                \be
                                \Sigma = U\rho^4  \Gamma_T \Gamma_* \left[\Gamma_\Upsilon + Z'
                                \Gamma_Z\right]\, vol_4(\xi)\, ,
                                \ee
                                and 
                                \be
                                \gamma^t\gamma^\rho = -\left[1+(Z')^2\right]^{-1}
                                \Gamma_T\left[ \Gamma_\Upsilon
                                + Z' \Gamma_Z\right].
                                \ee
                                Given also (\ref{dbilag}), we then deduce, after some algebra,
                                that\footnote{Here, and henceforth, we supress the tensor
                product
                                symbol.}
                                \be
                                \label{susycon}
                                \sqrt{1 + (Z')^2 -E^2} \ \epsilon =  \sigma_1
                                \Gamma_T\Gamma_*\Gamma_\Upsilon\ \epsilon
                                + i\sigma_2 \Gamma_*\left(E- Z' \sigma_3
                \Gamma_T\Gamma_Z\right)
                                \epsilon \, . 
                                \ee
                                Note that the function $U$ has cancelled, so the final result
                must be
                                the
                                same
                                as for flat space! 

                                The constraint (\ref{susycon}) must be satisfied for all
                $\rho$.
                                Because
                                $E$ and $Z'$ vanish asymptotically, as
                $\rho\rightarrow\infty$, we
                                deduce
                                from (\ref{susycon}) the `D5-constraint'
                                \be
                                \label{d5constraint}
                                \sigma_1\Gamma_T\Gamma_*\Gamma_\Upsilon\ \epsilon = \epsilon\,
                .
                                \ee
                                This is compatible with the D3-constraint (\ref{d3con}) and
                reduces
                                the
                                fraction of supersymmetry preserved to 1/4. Given that
                $\epsilon$ also
                                satisfies (\ref{d3con}), the supersymmetry preserving
                condition can
                                now
                                be reduced to
                                \be
                                \label{resid}
                                \left[\sqrt{1 + (Z')^2 -E^2} -1\right] \epsilon =
                                (E-Z')\Gamma_\Upsilon\Gamma_Z \epsilon\, .
                                \ee
                                In the cylindrical polar coordinates used here, $\epsilon =
                                M(\xi)\epsilon_0$
                                for constant spinor $\epsilon_0$ and matrix function $M$ of
                the
                                4-sphere
                                angles $\{\xi\}$. As $M$ does not commute with $\Gamma_\Upsilon$,
                the
                                equation  (\ref{resid}) can be satisfied for all $\{\xi\}$ if and
                only if
                                \be
                                E=Z'\, .
                                \ee
                                Thus, any static $SO(5)$-invariant D5-brane configuration with
                $E=Z'$
                                preserves 1/4 supersymmetry.

                                Using $E=Z'$ in the integrated Gauss law constraint
                (\ref{gl}), we
                                deduce
                                that
                                \be
                                \label{gl2}
                                U\rho^4 Z'  =  -{3\over2} L^4
                                \left[\theta - \sin\theta \cos\theta -{2\over3}\sin^3\theta
                                \cos\theta -\pi\nu\right] .
                                \ee
                                Given the form (\ref{Unow}) of the function $U$, and the
                relation
                                (\ref{ztheta}) between the functions
                                $Z$ and $\theta$, it can be shown \cite{CRS} that this is
                equivalent
                                to
                                \be
                                \label{zprime}
                                Z' = \left[ L^4\left(\theta -\sin\theta\cos\theta -
                                \pi\nu\right)/2\rho^3\right]'\, .
                                \ee
                                This is trivially integrated, and the result is the implicit
                equation
                                (\ref{finald5}) for $Z$  quoted in the introduction, which we
                may
                                write as
                                \be
                                \label{zeq}
                                Z= Z_\infty + {L^4\eta_\nu(\theta)\over 2\rho^3}
                                \ee
                                where
                                \be
                                \eta_\nu(\theta) = \theta -\sin\theta\cos\theta - \pi\nu\, .
                                \ee

                                \section{Energy and Force}
                                \label{energetics}

                                The interpretation of our results to follow will rely on a
                formula for
                                the D5-brane energy that we now derive. The effective D5-brane
                Hamiltonian
                                density is
                                \be
                                {\cal H} \equiv E {\partial {\cal L}\over \partial E} - {\cal
                L} =
                                 {8\pi^2\over 3} T_5\  {\rho^4U \left[1+(Z')^2\right]\over
                                \sqrt{1+(Z')^2-E^2}}\, .
                                \ee
                                Setting $E=Z'$ and using (\ref{gl2}), we find that
                                \be
                                \label{energy1}
                                {\cal H} = (8\pi^2 T_5/3) \rho^4U + Z'D_\nu(\theta)
                                \ee
                                for supersymmetric D5-branes, where
                                \be
                                \label{dtheta}
                                D_\nu = {NT_f\over \pi}\left[ \pi\nu -\theta +
                \sin\theta\cos\theta +
                                {2\over3}\sin^3\theta\cos\theta\right].
                                \ee
                                Here we have used the relation
                                \be
                                4\pi^3 T_5 L^4 = NT_f\, ,
                                \ee
                                which follows from (\ref{L}) and (\ref{t5}). The first term in
                                (\ref{energy1}) is the energy density
                due to the D5 surface
                tension. The
                                second term is due to the BI electric field, and its form
                shows that
                                $D_{\nu}$ can
                                be interpreted as a tension along the Z-axis, at least
                whenever it is
                                approximately constant. This observation is crucial to the
                                interpretation
                                of the 1/4 supersymmetric D5-brane configurations, on which we
                                elaborate in
                                the following sections.
                                 
                                The formula for the energy density (\ref{energy1}) can also be
                written
                                in
                                the
                                form \cite{CGMV}
                                \be
                                {\cal H} = {\cal H}_0 + \left[(2NT_f/3\pi)\rho\sin^4\theta +
                                ZD_\nu\right]'\, ,
                                \ee
                                where
                                \be
                                {\cal H}_0 = {8\pi^2 \over3} T_5 \rho^4
                                \ee
                                is the energy density for a flat vacuum D5-brane in flat
                space. The
                                total
                                energy
                                \be
                                H= \int_0^\infty d\rho\; {\cal H}
                                \ee
                                is of course infinite. One can subtract the infinite integral
                of
                                ${\cal
                                H}_0$, but the remainder is still infinite because of the term
                linear
                                in
                                $\rho$. 
                \FIGURE{
                \label{figure1}
                {\par\centering \includegraphics{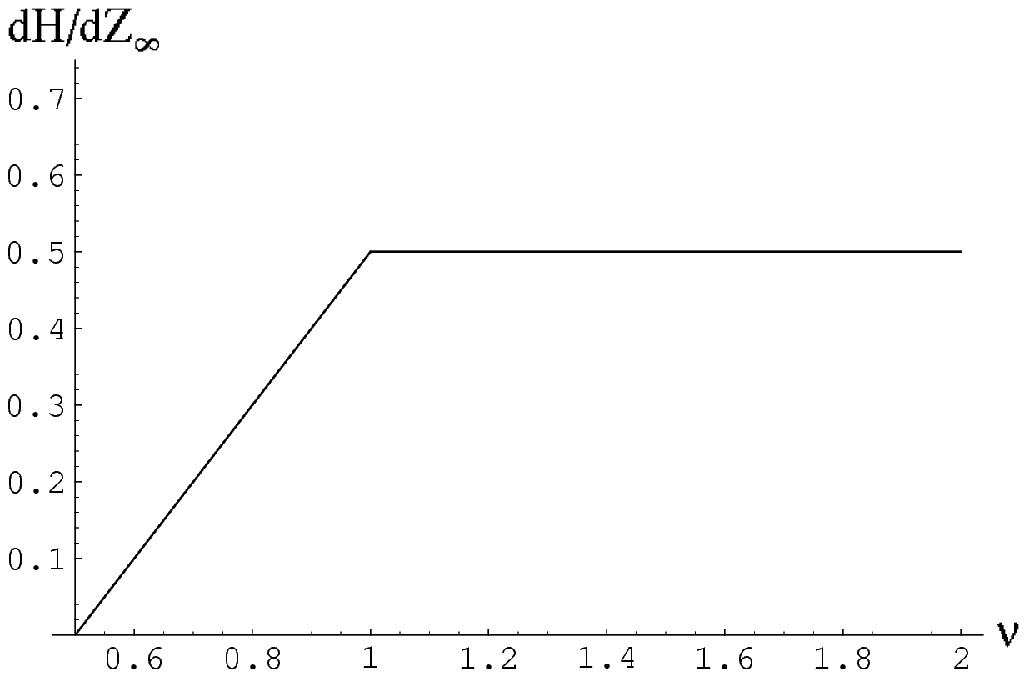} \par}
                \caption{The net force (in units $NT_f$) on the D5-brane as a function of
                $\nu$.}
                }
                                One can get a finite result by considering the derivative with
                                respect to $Z_\infty$. Noting that
                                \be
                                dZ/dZ_\infty = U^{-1}\, ,
                                \ee
                                one can show that 
                                \be
                                {d{\cal H} \over d Z_\infty} =  {2NT_f\over 3\pi
                L^4}\left[U^{-1}D_\nu\right]'.
                                \ee
                                Using (\ref{gl2}), and
                                integrating over $\rho$, we deduce that
                                \be
                                {dH \over d Z_\infty} =  {2NT_f\over 3\pi L^4}\left[\rho^4
                                Z'\right]_0^\infty\, .
                                \ee
                                One sees immediately from (\ref{zeq}) that
                                \be\label{asymp}
                                \rho^4 Z' \rightarrow {3\pi L^4\over 2} 
                                \left(\nu-{1\over2}\right)
                                \ee
                                as $\rho\rightarrow\infty$. An analysis of the behaviour of
                $Z$ as
                                $\rho\rightarrow0$ will be considered in more detail in the
                                following sections. We will see that, as $\rho\rightarrow 0$,
                                \be
                                \rho^4 Z' \rightarrow \left \{ \begin{array}{cl} 0 &
                                \quad \textrm{for } 1/2\leq \nu \leq 1 \\ {3\pi L^4 \over 2}
                                (\nu-1) & \quad \textrm{for } \nu > 1
                \end{array} \right. .
                                \ee
                                This yields the result
                                \be
                                {dH \over d Z_\infty} = \left \{ \begin{array}{cl}
                                NT_f(\nu-{1\over2}) & \quad \textrm{for } 1/2\leq \nu <1 \\
                                {1\over2}NT_f & \quad \textrm{for } \nu \geq 1
                \end{array}\right. .
                                \ee
                                Thus, the net force vanishes {\sl only} for
                                $\nu=1/2$. Fig. \ref{figure1} shows a plot of the
                                force $dH/dZ_\infty$ as a function of $\nu$.

                As $E=Z'$, the asymptotic behaviour of $Z'$ given in (\ref{asymp})
                shows that the BI electric charge as measured by the electric flux 
                at infinity is proportional to $(\nu-1/2)$. This vanishes for $\nu=1/2$,
                so in this case there is no obstruction to a compactification of the
                D5-brane. We could compactify on a torus and then T-dualize to the
                D0-D8 system, for which the force is known to vanish \cite{BGL}, so we
                should find a vanishing force on the D5-brane when $\nu=1/2$, and we
                do. When $\nu\ne1/2$ no such argument applies and, as a
                compactification is not possible, the total mass of the D5-brane is
                necessarily infinite. As an infinite mass cannot be moved by a finite
                force, a non-zero force is compatible with the fact that the D5-brane
                is static. As we have seen, it is also compatible with supersymmetry.

                                \section{String creation and the s-rule}

                                The key equation (\ref{zeq}) is equivalent to
                                \be
                                \tilde Z= \tilde Z_\infty +
                {L^4\eta_{(1-\nu)}(\tilde\theta)\over
                                2\rho^3}
                                \ee
                                where
                                \be 
                                 \tilde Z = -Z\, ,\qquad \tilde Z_\infty = -Z_\infty \,
                ,\qquad
                                \tilde \theta = \arctan (\rho/\tilde Z)\, .
                                \ee
                                This shows that there is no loss of generality in restricting
                $\nu$ in
                                (\ref{zeq}) to the range $\nu\ge {1\over2}$, as claimed in the
                                introduction,
                                as long as we allow $Z_\infty$ to be either positive or
                negative. When
                                $Z_\infty>0$ and $\nu\le1$, the constant $\nu N$ has the
                                interpretation as
                                the
                                number of strings connecting the D5-brane to the N D3-branes
                                (although,
                                strictly speaking, this interpretation makes sense only for
                $\nu=1$).
                                But
                                when
                                $Z_\infty<0$ one finds that the D5-brane is connected to the
                D3-branes
                                by
                                $(\nu-1)N$ anti-strings, and this is responsible for the HW
                effect.
                                The
                                interpretation is simplest for $\nu=1$, for which, following
                                \cite{CRS}, we
                                present plots of $Z(\rho)$ for various values of
                                $Z_\infty$, see fig. \ref{figure2}. These
                                plots
                                clearly exhibit the mechanism underlying the effect;
                essentially, the
                                $\rho
                                <L$ region of the D5-brane remains trapped on a 5-sphere
                surrounding
                                the
                                D3-branes as the D5-brane is pulled through them, but this
                wrapped
                                D5-brane
                                remains connected to the asymptotic D5-brane by a tube of $S^4$
                cross
                                section
                                that can be interpreted as $N$ strings. This explains the HW
                effect
                                for
                                $\nu=1$; a different explanation is needed for $\nu>1$ and
                this will
                                be
                                provided below.
                \FIGURE{\label{figure2}
                {\par\centering \includegraphics{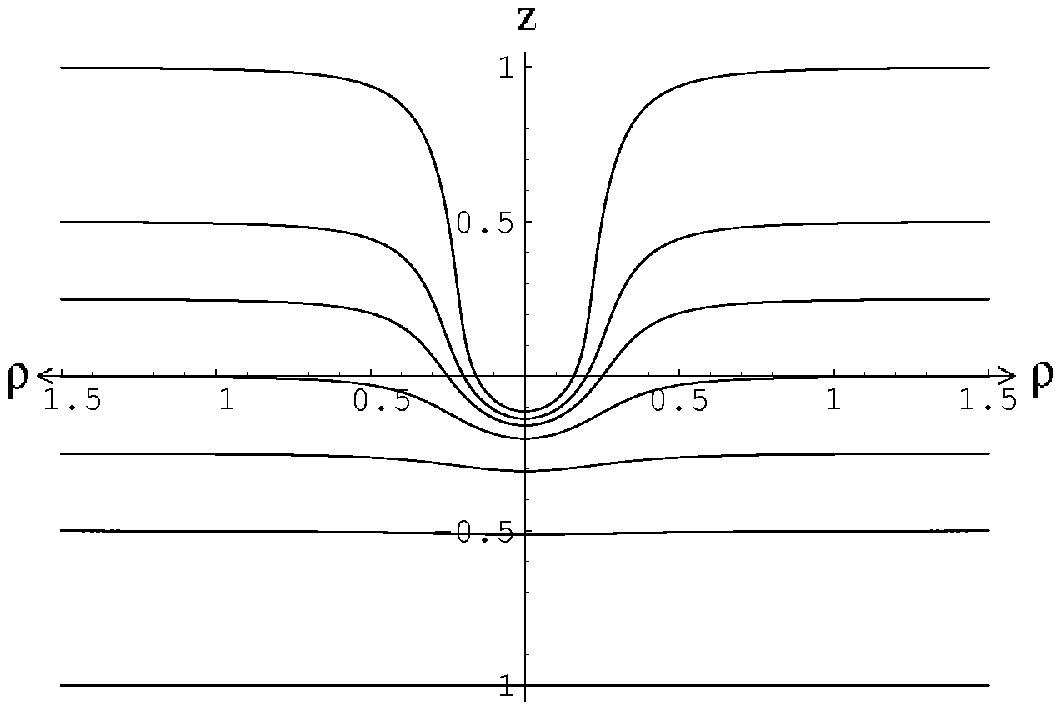} \par}
                \caption{\protect\( \nu =1\protect \): \protect\( Z(\rho )\protect \) (and its
                mirror
                image) for various values of \protect\( Z_{\infty }\in [-1,1]\protect \).}
                }
                               
                 When $Z\ll Z_\infty$ we can approximate (\ref{zeq}) by
                                \be
                                \eta_\nu(\theta) = -{2Z_\infty \over L^4} \rho^3\, .
                                \ee
                                Defining a new independent variable $r$ (radial distance
                                in the $\bE^6$ transverse to the D3-branes) by
                                \be
                                \rho = r \sin\theta
                                \ee
                                we can rewrite the above equation as
                                \be
                                \label{nearh}
                                r^3  = - {L^4 \over 2 Z_\infty} \left({\eta_\nu(\theta) \over
                                \sin^3\theta}\right) \, ,
                                \ee
                                which is equivalent to the result of \cite{CGS} for a D5-brane
                                in $adS_5 \times S^5$. For $\nu=1$ there is a minimum value
                $r_{min}$
                                of
                                $r$,
                                with $r=r_{min}$ for $\theta =\pi$, and
                                \be
                                Z_\infty r^3_{min} \sim L^4\, .
                                \ee
                                Clearly, the near-horizon approximation to the D3 geometry is
                valid
                                only if
                                $r_{min} \ll L$, but this condition is satisfied only if
                                \be\label{validity}
                                Z_\infty \gg L .
                                \ee
                                In other words, the `near horizon' D5 probe geometry of
                (\ref{nearh}),
                                which
                                describes a D5-brane wrapped on an $S^5$ in $adS_5\times S^5$
                with N
                                strings
                                attached \cite{CGS}, can be interpreted as a D5-brane
                surrounding N
                                D3-branes at a distance $r_{min}$ only if the N strings
                connect to a
                                {\sl
                                distant} planar D5-brane. However, string creation occurs as
                                $Z_\infty$
                                passes through zero, at which point the condition
                (\ref{validity})
                                must fail. We conclude that the near-horizon result
                                (\ref{nearh}) is actually {\sl not} relevant to a worldvolume
                                description of
                                the HW effect, although it is still a useful tool in the
                analysis of
                                (\ref{zeq}).

                                We have been viewing $Z$ and $\theta$ as functions of $\rho$
                that are
                                related
                                by (\ref{ztheta}).  However one can view  (\ref{ztheta}) as a
                single
                                relation
                                between three variables $(Z,\theta,\rho)$, any one of which
                may be
                                chosen as
                                the independent variable. For some purposes it is convenient
                to choose
                                $\theta$ as the independent variable, in which case
                (\ref{ztheta}) can
                                be
                                interpreted as defining the function
                                \be
                                \rho(\theta) = Z(\theta)\tan\theta\, .
                                \ee
                                Using this in (\ref{zeq}) we deduce that the function
                $Z(\theta)$ is
                                given
                                implicitly by the relation
                                \be
                                Z= Z_\infty +  {L^4 \eta_\nu(\theta) \cot^3\theta \over 2
                Z^3}\, .
                                \ee
                                The function $Z(\theta)$ is not necessarily single-valued but
                there is
                                always a branch near $\theta=\pi/2$ for which $Z\approx
                Z_\infty$.
                                The induced metric on this branch approaches the flat
                Minkowski metric
                                as
                                $\theta\rightarrow \pi/2$. Plots of $Z(\theta)$ are
                                shown in fig. \ref{figure3} for $\nu=1$ and $Z_\infty >0$ or
                                $Z_\infty <0$, respectively. 
                \FIGURE{\label{figure3}
                {\par\centering \includegraphics{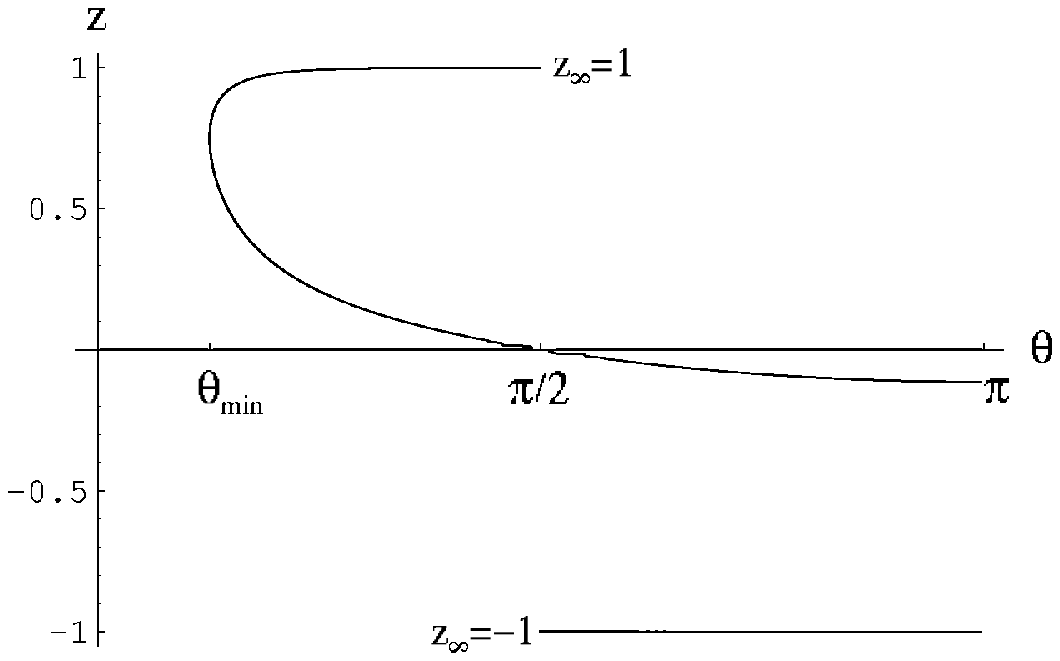} \par}
                \caption{\protect\( \nu =1\protect \): \protect\( Z(\theta )\protect
                \) for \protect\( Z_{\infty }=-1\protect \) 
                and \protect\( Z_{\infty }=1\protect \). }
                }
                For
                                $Z_\infty <0$,
                                one sees, as expected, that $Z$ remains everywhere close to
                                $Z_\infty$.
                                For $Z_\infty >0$, however, $Z(\theta)$ is doubled-valued at
                $\theta=\pi/2$,
                                and vanishes on the second branch. On this branch, $Z$ has a
                minimum
                                at
                                $\theta=\pi$, where its value is small and negative. The two
                branches
                                are
                                connected by a region in which $\theta$ approaches a small
                minimum
                                value $\theta_{min} \sim (L/Z_\infty)^{4/3}$ at
                $Z(\theta_{min})=3Z_\infty /4$.
                                In this region the
                                D5 geometry is that of a thin tube along the Z-axis with a
                                variable-radius
                                4-sphere as its cross section, as shown in fig. \ref{figure4}.
                                There is a similar tube for all $\nu\ge
                                1/2$ (given $Z_\infty \gg L$). From (\ref{energy1}) we see
                that the
                                energy
                                per unit length of this tube is proportional to
                                $D_\nu$, and from (\ref{dtheta}) we have, near $\theta=0$,
                                \be
                                \label{dzero}
                                D_\nu(\theta) = NT_f\nu + {\cal O}(\theta^5)\, ,
                                \ee
                                The tube's tension is therefore $N\nu$ times the IIB string
                tension.
                                For
                                $\nu=1$ we can therefore interpret the tube as N IIB strings
                stretched
                                between the D5-brane and the N D3-branes.
                \FIGURE{\label{figure4}
                {\par\centering \includegraphics{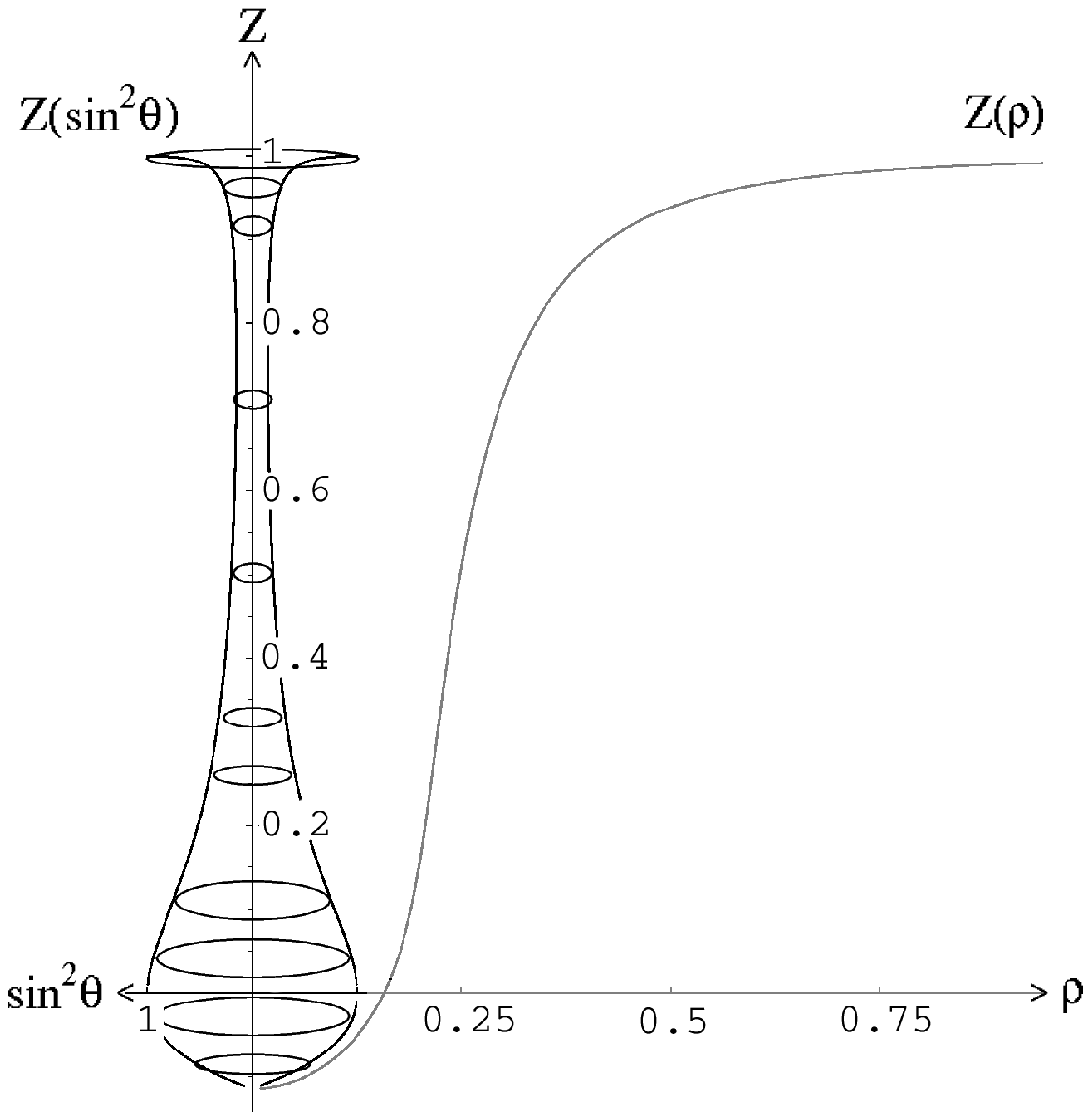} \par}
                \caption{\protect\( \nu =1\protect \): \protect\( Z(\rho )\protect \) and the
                string
                interpretation. \protect\( Z\protect \) as a function of the \protect\(
                S^{4}\protect \)-size
                \protect\( \sin ^{2}\theta \protect \). }
                }

                                All previous analyses of the D5-brane worldvolume dynamics
                have been
                                subject
                                to the restriction $0\le \nu \le 1$. Given that we may choose
                $\nu\ge
                                1/2$
                                without loss of generality, this means in effect that only the
                cases
                                with
                                \be
                                {1\over2} \le \nu \le 1
                                \ee
                                have been considered previously. The $\nu=1$ case allows the
                simplest
                                realization of the HW effect, as just reviewed. We have
                nothing new to
                                say
                                about the $\nu <1$ cases, except for the special case of
                $\nu=1/2$
                                which
                                will
                                be dealt with in the following section. This leaves the cases
                for
                                which
                                \be
                                \nu >1\, .
                                \ee
                                For simplicity of presentation we shall assume that $\nu$ is
                an
                                integer;
                                this
                                means that (when $Z_\infty>0$) we have a D5-brane in the D3
                background
                                with
                                $\nu N$ strings attached to it. Our initial ansatz imposed an
                $SO(5)$
                                symmetry which forces all these attached strings to lie along
                the axis
                                ($\theta=0,\pi$) separating the D5-brane from the D3-branes.
                Although
                                one
                                should take $N$ large to justify the supergravity
                approximation, the
                                results
                                make formal sense for any $N$ and it will be convenient to
                discuss the
                                $N=1$
                                case. We shall also assume, at least initially, that
                $Z_\infty>0$. For
                                $\nu=1$ we then have a D5-brane geometry that can be
                interpreted as a
                                D5-brane connected to a D3-brane by a single string. When
                $\nu>1$ we
                                must
                                have $\nu$ strings that leave the D5-brane in the direction of
                the
                                D3-brane
                                (because any leaving in the opposite direction would be
                                supersymmetry-breaking anti-strings). However, according to
                the
                                s-rule, only
                                one of these $\nu$ strings can end on the D3-brane, so {\it
                the other
                                $(\nu-1)$ strings must pass through the D3-brane, without
                ending on
                                it}.
                \FIGURE{\label{figure5}
                {\par\centering \includegraphics{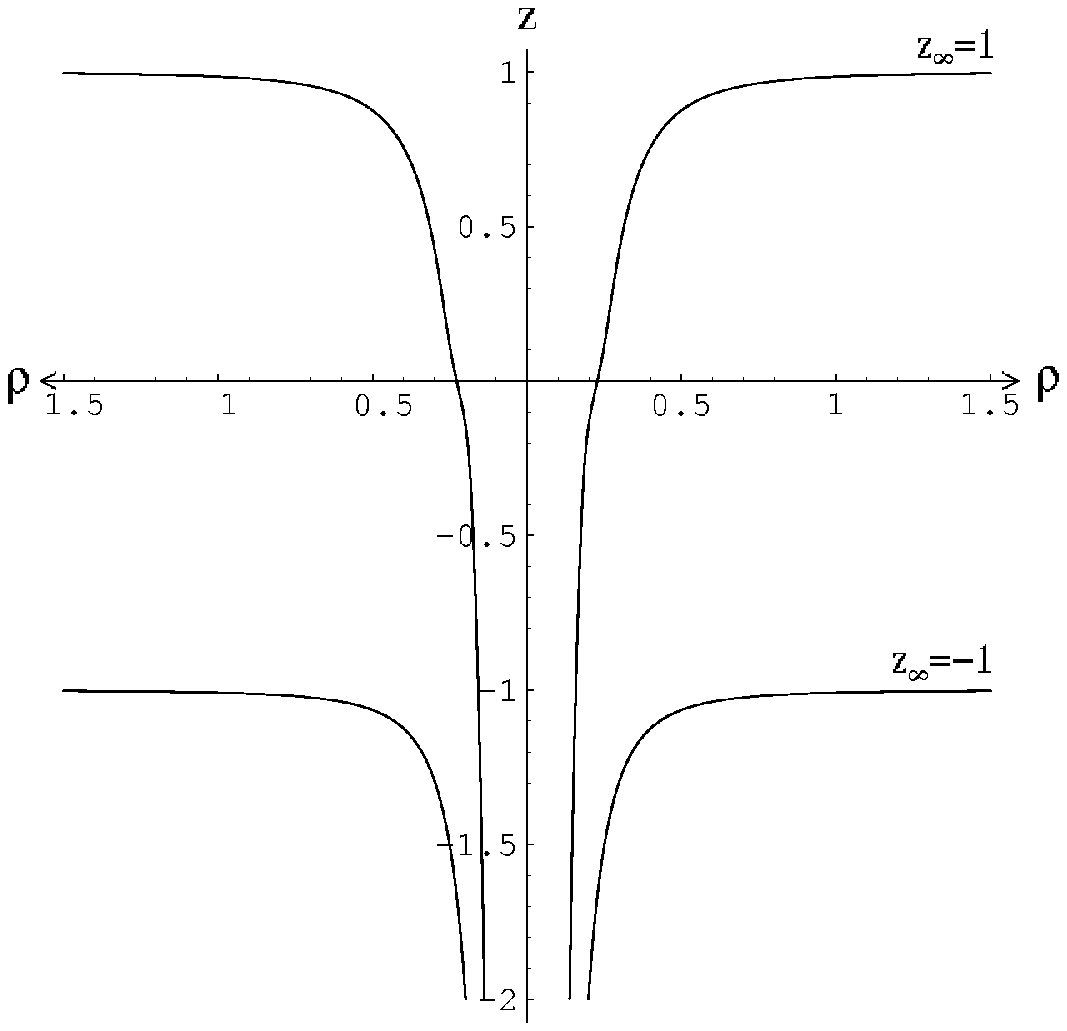} \par}
                \caption{\protect\( \nu =2\protect \): \protect\( Z(\rho )\protect \) (and its
                mirror
                image) for \protect\( Z_{\infty }=-1\protect \) and \protect\( Z_{\infty
                }=1\protect \).}
                }
                                This conclusion may be verified qualitatively by inspection of
                the
                                plot of
                                the
                                function $Z(\rho)$ for $\nu=2$ (with $Z_\infty>0$),
                                see fig. \ref{figure5}; comparison with
                                the plot
                                for $\nu=1$ (fig. \ref{figure2}) shows that at least one
                string now passes {\sl through}
                                the D3-brane. It might appear that all
                pass through the D3-brane but a closer analysis shows
                that this interpretation is not correct. Consider the
                plot of the function $Z(\theta)$ shown in
                fig. \ref{figure6} for $\nu=2$.
                \FIGURE{\label{figure6}
                {\par\centering \includegraphics{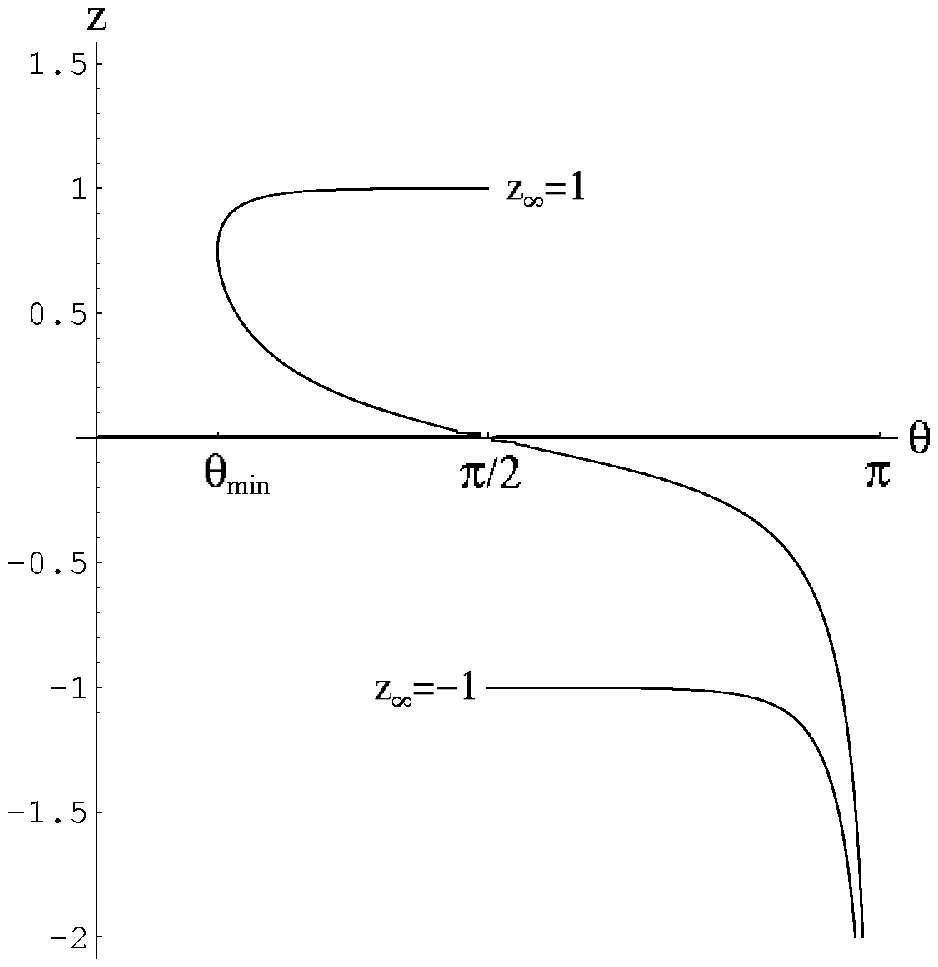} \par}
                \caption{\protect\( \nu =2\protect \): \protect\( Z(\theta )\protect \) for
                \protect\( Z_{\infty }=-1\protect \) and \protect\( Z_{\infty }=1\protect \).
                }
                }
                                Here we see that in addition to the asymptotic region as
                                $\theta\rightarrow
                                \pi/2$ there is now {\sl another asymptotic region} as
                                $\theta\rightarrow
                                \pi$. The geometry is that of an infinite BIon `spike'
                                \cite{CM,GWG} along the
                                Z-axis with
                                cross section $S^4$. The 4-sphere has an ever-decreasing
                                radius, but
                                the energy per unit length is again proportional to $D_\nu$.
                As
                                \be
                                D_\nu(\pi) = NT_f(\nu -1) \, ,
                                \ee
                                we may interpret this spike as $\nu-1$ strings which have
                passed
                                through
                                the D3-brane without ending on it; see fig. 7.
                \begin{figure}[h]\label{figure7}
                {\par\centering \includegraphics{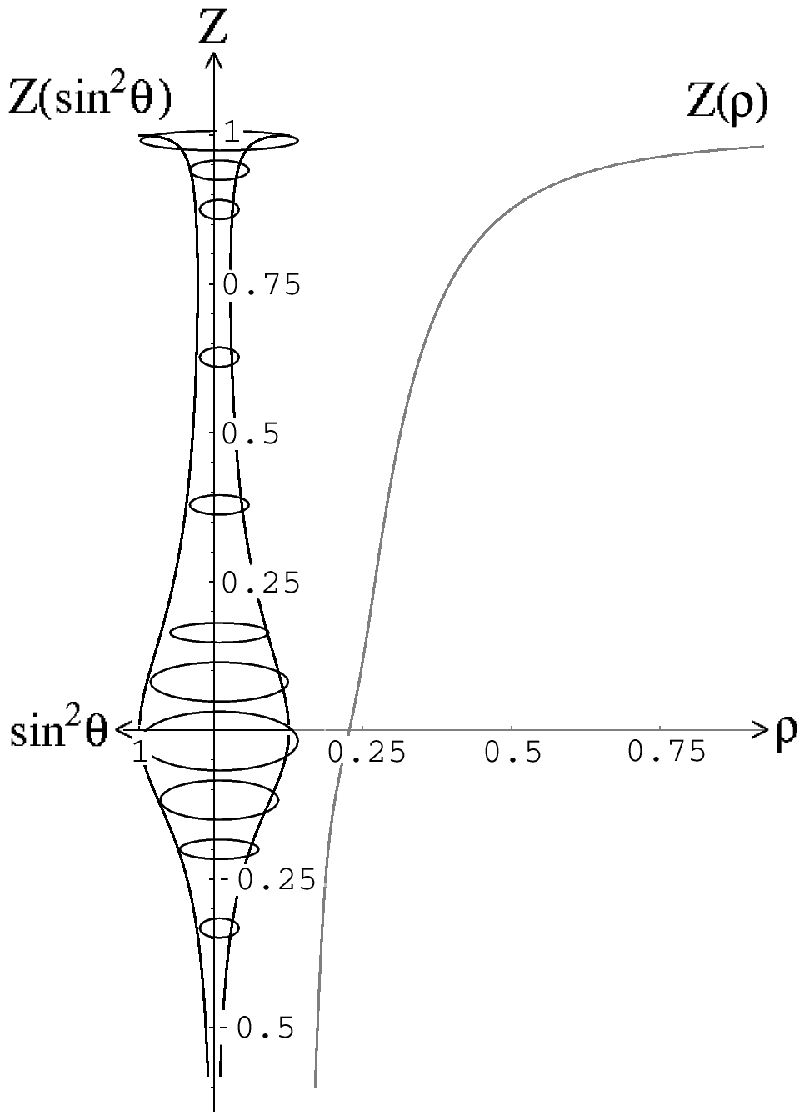} \par}
                \caption{\protect\( \nu =2\protect \): \protect\( Z(\rho )\protect \) and the
                string
                interpretation. \protect\( Z\protect \) as a function of the \protect\(
                S^{4}\protect \)-size
                \protect\( \sin ^{2}\theta \protect \). }
                \end{figure}
                                As we started with $\nu$ strings we must
                                conclude, {\sl despite the fact that the D5-brane does not
                cross the D3
                                horizon}, that one string has ended on the D3-brane.

                                The above discussion was for positive $Z_\infty$ and
                                $\nu>1$. Turn now to the plot of the function
                                $Z(\rho)$ for
                                negative
                                $Z_\infty$ in fig. \ref{figure5}. For $Z_\infty\gg L$
                                the
                                D5-brane is always far from the D3-branes and the D5 probe
                geometry so
                                there
                                are now no strings connecting the D5-brane to the D3-brane.
                 From the
                                plot of
                                $Z(\theta)$ in fig. \ref{figure6} one sees that there is again
                a second asymptotic
                                region
                                as $\theta\rightarrow \pi$, but this corresponds to $(\nu-1)$
                                strings that leave the D5-brane in the {\sl opposite
                                direction to
                                the D3-branes}. As $Z_\infty$ is taken from positive to
                negative values
                                the
                                one
                                string stretched between the D5 and the D3 is therefore
                destroyed,
                                thus
                                realizing the HW effect for $\nu>1$.  In no case is the D5
                connected
                                to the
                                D3 by more than one string, thus confirming the s-rule.

                                \section{Vacuum polarization and vacuum interpolation}

                                For $1/2 \le \nu < 1$ and $Z_\infty \gg L$ there is always a
                                `near-horizon'
                                branch of the function $Z(\theta)$ for which we may use the
                                near-horizon
                                relation (\ref{nearh}). This shows that there exists a maximum
                value
                                $\theta_{max}^{(\nu)}$ of $\theta$ (with $\pi/2 \le \theta
                                <\pi$, as illustrated in fig. \ref{figure8}) and that
                                \be
                                \theta \rightarrow \theta_{max}^{(\nu)} \Rightarrow r\rightarrow 0\, ,
                                \ee
                                where $r^2=\rho^2 + Z^2$.
                                This means that the D5-brane crosses the D3 horizon when $\nu$
                lies in
                                the
                                range $1/2 \le \nu < 1$, in contrast to its
                                behaviour for $\nu\ge 1$.  We shall concentrate on the
                $\nu=1/2$ case,
                                for
                                which $\theta_{max}^{(0.5)} = \pi/2$. The function $Z(\theta)$
                                for this case is shown in fig.
                                \ref{figure8} for $Z_\infty=1$; 
                                note that  there are {\sl two} branches of the function at
                                $\theta=\pi/2$, the near-horizon branch just discussed and an
                                asymptotic
                                branch on which the induced geometry is Minkowski.
                \FIGURE{\label{figure8}
                {\par\centering \includegraphics{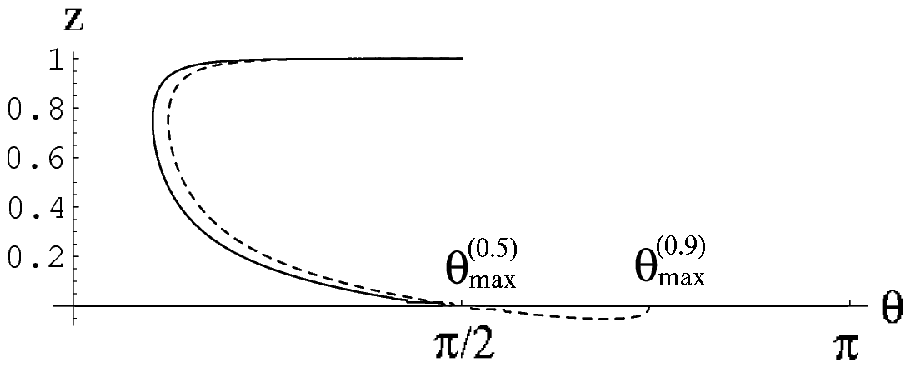} \par}
                \caption{\protect\( \nu =0.5\protect \) and \protect\( \nu
                =0.9\protect \) (dashed): \protect\( Z(\theta )\protect \)
                for \protect\( Z_{\infty }=1\protect \). }
                }

                                {}From the formula (\ref{dtheta}) we see that
                                \be
                                D_\nu(\pi/2) = N\left(\nu -{1\over2}\right)T_f
                                \ee
                                and hence that $D_\nu \rightarrow 0$ asymptotically only if
                $\nu=1/2$. In
                                addition, when $\nu=1/2$ we have $D_\nu\rightarrow 0$ on the
                near-horizon
                                branch
                                too! This means that $D_\nu(\theta)$ is non-vanishing only in
                the
                                tubular region connecting the near-horizon and asymptotic
                branches of
                                the
                                D5-brane; in other words, {\it the D3-brane polarizes the
                electrically
                                neutral D5-brane}, separating
                                equal but opposite amounts of BI electric charge across the
                tubular
                                region.
                                The resulting BI electric field can be interpreted as $N$
                                half-strings.
                                This observation provides a simple
                                explanation of the HW effect for the $\nu=1/2$ case. As
                $Z_\infty$ is
                                reduced
                                from its large positive value, the polarized region shrinks
                until, at
                                $Z_\infty=0$, it vanishes (as we confirm below). As $Z_\infty$
                                continues to
                                decrease to large negative values the polarization reappears
                but now
                                with
                                the
                                opposite orientation, thus creating $N$ anti-half-strings. The
                net
                                effect is
                                to destroy $N$ strings (or create them, depending on the
                initial
                                choice of
                                space orientation).

                                We shall now examine the D5-brane geometry for
                                $\nu=1/2$ in more detail.
                                Setting $\nu=1/2$ in (\ref{zeq}) we have
                                \be
                                \label{nuhalf}
                                Z = Z_\infty + {L^4 \over 2\rho^3}
                                \left[\theta - {\pi\over 2} - \sin\theta\cos\theta \right]\, .
                                \ee
                                This equation determines a family of functions $Z(\rho)$, or
                                equivalently
                                $\theta(\rho)$ with $\tan\theta = \rho/Z$, depending on the
                parameter
                                $Z_\infty$. As long as $Z_\infty \ne 0$, there is a branch of
                the
                                function
                                with $Z\approx 0$ that is described in the limit
                $\rho\rightarrow 0$
                                by the
                                near-horizon approximation, which yields
                                \be
                                \theta \sim {\pi\over 2} -
                                {Z_\infty\over L^4}\rho^3 .
                                \ee
                                It follows from this that $\rho^4Z'\rightarrow 0$ as
                $\rho\rightarrow
                                0$, as
                                claimed in section \ref{energetics}.

                \FIGURE{\label{figure9}
                {\par\centering \includegraphics{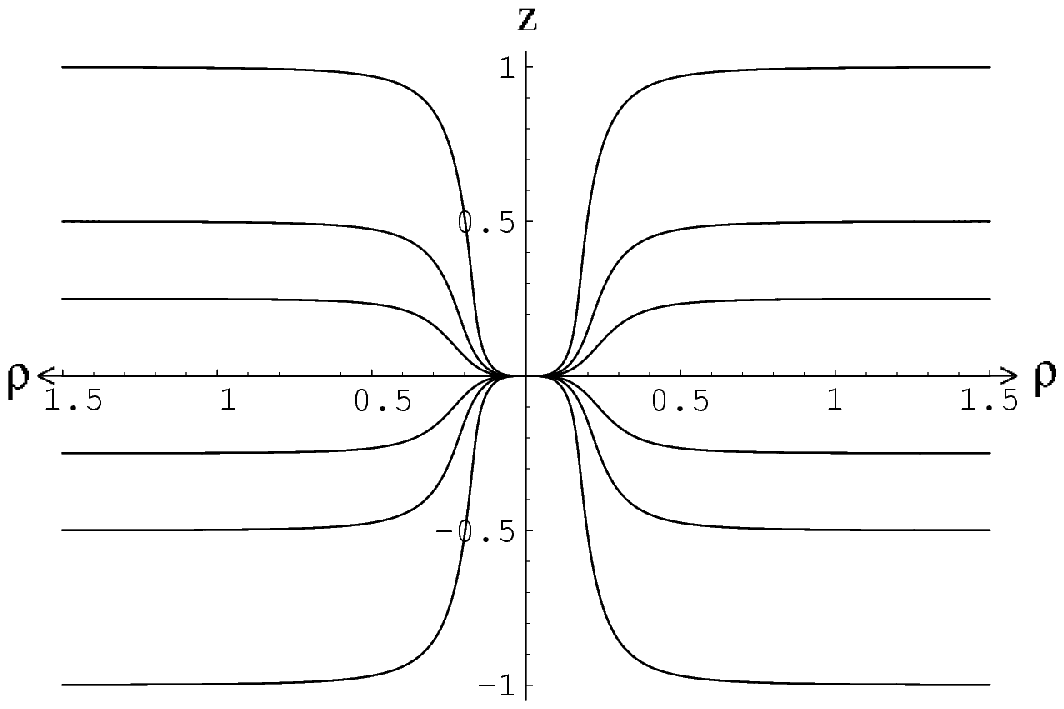} \par}
                \caption{\protect\( \nu =0.5\protect \): \protect\( Z(\rho )\protect \) (and
                its mirror
                image) for various values of \protect\( Z_{\infty }\in [-1,1]\protect \).}
                }
                                It also follows that $Z\rightarrow 0$ as $\rho\rightarrow 0$.
                As can
                                be seen
                                from fig. \ref{figure9} the region in which $Z\approx 0$ grows
                as
                                $Z_\infty
                                \rightarrow 0$. This suggests that $Z\equiv 0$ when
                $Z_\infty=0$. To
                                verify
                                this we must return to (\ref{nuhalf}) and set $Z_\infty=0$.
                The
                                resulting
                                equation indeed has the solution
                                \be
                                Z_\infty \equiv 0\, .
                                \ee
                                To see that this is the {\sl only} solution we note
                                that if $Z_\infty=0$ but $Z\ne0$
                                then
                                (\ref{nuhalf}) implies that
                                \be
                                Z^4 = -{1\over2}L^4 \cot^3\theta\left[(\pi/2) -\theta +
                                \sin\theta\cos\theta\right] \le 0\, ,
                                \ee
                                which is impossible.

                                Let us now return to the generic $\nu=1/2$ case, for which
                                $Z\not\equiv 0$.
                                Because $D_\nu(\pi/2)=0$, we have
                                $D_\nu\rightarrow 0$ as $\rho\rightarrow 0$. Using the
                                near-horizon
                                approximation for $U$ we compute the induced metric in this
                limit to
                                be
                                \be
                                \label{adSS}
                                ds^2 = - {\rho^2\over L^2} dt^2 + L^2 {d\rho^2\over \rho^2} +
                L^2
                                d\Omega_4^2\, .
                                \ee
                                This is an $adS_2\times S^4$ embedded in the near-horizon
                $adS_5\times
                                S^5$
                                background with $E=Z'=0$. We thus have a {\sl worldvolume}
                analogue of
                                the
                                interpolation property of the D3-brane background \cite{GT}.
                Recall
                                that in
                                this case both the asymptotic Minkowski vacuum and the
                near-horizon
                                $adS_5
                                \times S^5$ vacuum are maximally supersymmetric, with twice
                the number
                                of
                                supersymmetries of the full D3 supergravity solution. We shall
                now
                                show that
                                the interpolating D5-brane has the same property by
                demonstrating that
                                its
                                $adS_2\times S^4$ vacuum has enhanced supersymmetry.

                                Our result of section \ref{d5revisit} for the
                                supersymmetry preserving constraint arising from the presence
                of the
                                D5-brane
                                is equivalent to the condition
                                \be
                                \label{d5conchi}
                                \sigma_1\Gamma_T\Gamma_*\Gamma_\Upsilon
                                \; \chi = \chi
                                \ee
                                on background Killing spinors $\chi$. For the full D3
                background these
                                Killing spinors all take the form (\ref{chiform}) but for the
                                $adS_5\times S^5$ background there are additional Killing
                                spinors of the form \cite{LPT}
                                \be
                                \label{addspin}
                                \chi = \left(\Upsilon^2 +Z^2\right)^{-3/4} \left[
                                \Upsilon\Gamma_\Upsilon +
                                Z\Gamma_Z - (\Upsilon^2 +Z^2)\left( T\Gamma_T + \vec X \cdot
                                \vec\Gamma_X\right)\right]\eta
                                \ee
                                where $\eta$ is a covariantly constant spinor in the IIB
                Minkowski
                                vacuum subject to the $\overline{D3}$ constraint
                                \be
                                \label{d3bar}
                                i\sigma_2 \Gamma_Z\Gamma_\Upsilon\Gamma_*\;\eta = -\eta\, .
                                \ee
                                For our gauge choice and $\vec X \equiv 0$ ansatz we have
                                \be
                                \chi = \left[r^{-{3\over2}}\rho\Gamma_\Upsilon - r^{1\over2}t
                \Gamma_T
                                +
                                r^{-{3\over2}}Z\Gamma_Z\right]\eta\, .
                                \ee
                                Unless $Z$ vanishes identically, the constraint
                (\ref{d5conchi})
                                implies
                                that
                                $\eta=0$. However, when $Z\equiv0$ we find instead that $\eta$
                must
                                satisfy
                                \be
                                \sigma_1\Gamma_T\Gamma_*\Gamma_\Upsilon\;
                                \eta = -\eta\, .
                                \ee
                                This is compatible with (\ref{d3bar}) and together these
                constraints
                                imply that $\eta$ has eight independent components. The spinor
                                $\epsilon$ also has eight independent
                                components so the number of supersymmetries is doubled when
                                $Z\equiv0$. 

                                The isometry group of the embedding of $adS_2\times S^4$ in
                                $adS_5\times
                                S^5$ is a cover of $Sl(2;\bR)\times Sp_2$, which should
                therefore be a
                                subgroup of the isometry supergroup, with the 16 charges in
                two $({\bf
                                2},{\bf 4})$ representations of  $Sl(2;\bR)\times Sp_2$. This
                                supergroup
                                must
                                also be a subgroup of the $SU(2,2|4)$ isometry supergroup of
                the
                                background.
                                The only candidate \cite{CGMV} is the supergroup
                                \be
                                OSp(4^*|4) \supset Spin^*(4) \times USp(4) \cong
                Sl(2;\bR)\times
                                SU(2)\times Sp_2
                                \ee
                                for which the 16 supercharges are in the $({\bf 2},{\bf
                2},{\bf 4})$
                                irreducible representation of $Sl(2;\bR)\times SU(2)\times
                Sp_2$. The
                                additional $SU(2)$ factor has a natural interpretation as the
                rotation
                                group
                                acting on $\vec X$.

                                \section{Discussion}

                                The end result of a series of previous papers devoted to the
                                D5 worldvolume
                                interpretation of the Hanany-Witten effect is the implicit
                formula
                                (\ref{finald5}) that determines the geometry of a D5-brane in
                the D3
                                background. Here we have given a much simplified derivation of
                this
                                formula
                                from the condition for preservation of supersymmetry. The
                                simplification is
                                largely due to a better gauge choice, one that is adapted to
                the full
                                D3
                                geometry rather than its near-horizon limit. Indeed, one of
                the
                                lessons of
                                our work that was not fully appreciated previously is that the
                full D3
                                geometry is needed for a worldvolume interpretation of the HW
                effect.
                                Our new perspective on this problem has the virtue that it
                allows a
                                straightforward physical interpretation of the {\sl general}
                                supersymmetric $SO(5)$-invariant
                                'baryonic'
                                D5-brane in a D3 background, with arbitrary numbers of
                                attached IIB strings (corresponding to arbitrary
                                $\nu$); in agreement with the `s-rule', we
                                have
                                found that at most one of these strings may end on the
                D3-brane. This
                                provides a classical
                                interpretation, in the spirit of \cite{BG}, for what is
                usually
                                interpreted
                                as a quantum effect in IIB string theory due to the Pauli
                exclusion
                                principle. 

                                In the special case corresponding to $\nu=1/2$, for which
                                the D5-brane has no net BI electric charge, the
                                D5-brane
                                must cross the D3 horizon. The near horizon solution was found
                already
                                in
                                \cite{CGS} but its adS geometry was not previously apreciated,
                nor the
                                fact
                                that this solution has enhanced supersymmetry. Its
                $OSp(4^*|4)$
                                isometry
                                supergroup was discussed previously \cite{CGMV}, but without
                reference
                                to
                                the
                                solution that actually exhibits it. Here we have presented
                what we
                                hope is a
                                complete account of this special D5-brane embedding in
                $adS_5\times
                                S^5$.
                                In fact, this special case was the starting point of our work;
                it is
                                not
                                difficult to see that an $adS_2\times S^4$ embedding must
                exist, and
                                it
                                was  recently argued that it should have an interpretation,
                via the
                                adS/dCFT correspondence \cite{KR,dWFO}, as a point defect in
                ${\cal
                                N}=4$
                                SYM-theory \cite{ST,NMT}. If so, one might expect this defect
                to
                                support an
                                ${\cal N}=8$ supersymmetric conformal quantum mechanics
                (SCQM), and
                                one
                                might
                                expect supersymmetric deformations of the $adS_2\times S^4$
                D5-brane
                                to
                                correspond to non-conformal perturbations of this SCQM.
                However, the
                                supersymmetric deformations are the D5-brane geometries with
                                $Z_\infty\ne0$.
                                These asymptote to $adS_2\times S^4$ as $\rho\rightarrow 0$,
                which
                                corresponds to the {\sl IR limit} of the putative SCQM; there
                seems to
                                be no
                                supersymmetric deformation that is asymptotic to $adS_2\times
                S^4$ in
                                what
                                would be the UV limit.

                                Our improved analysis of the D5-brane energy allowed
                us to compute the force on the D5-brane as a function of the parameter
                $\nu$. This force need not vanish, despite supersymmetry, because a
                finite force cannot move an infinitely massive object. Naively, one
                might have expected the force for $\nu=1/2$ to be $(1/2)NT_f$ because
                in this case the D5-brane is connected to the D3-branes by N half-strings (at
                least
                if $Z_\infty\ne0$). In fact, the force vanishes when $\nu=1/2$, as it
                must in order to avoid contradiction with previous results for the
                T-dual D0-D8 system. Given this, it is understandable that the
                non-zero force for $\nu=1$ is $(1/2)NT_f$ rather than $NT_f$ (as one
                might naively have expected). The fact that the force remains at this
                value for all $\nu>1$ is a reflection of the fact that the addition of
                more strings has no effect on the D3-D5 dynamics; they pass straight
                through the D3-brane, as required by the s-rule.

                                There are various dual manifestations of the HW effect. One is
                the
                                creation
                                of M2-branes when two `linked' M5-branes cross \cite{HW,dA}.
                As in the
                                D5-D3
                                case, the M5-worldvolume geometry is determined in terms of a
                function
                                $Z(\rho)$ that gives the distance on the axis separating the
                M5-probe
                                from the background in terms of radial distance on the probe.
                Given
                                that the
                                M5-branes become D4-branes when reduced on their common
                direction, the
                                same
                                results should be obtained by considering a IIA D4-brane in a
                D4
                                background
                                (although justification of the supergravity approximation
                entails a
                                return
                                to
                                the M-theory description). This is indeed the case; the
                equation for
                                $Z(\rho)$ was found by energy minimization of a D4 probe in
                \cite{CRS}
                                and
                                from supersymmetry preservation of an M5-brane in \cite{GRST}.
                The
                                solution
                                of this differential equation is given implicitly by the
                algebraic
                                relation\footnote{This corrects a sign in eq. (3.58) of
                \cite{GRST},
                                which
                                was stated there as having been transcribed from \cite{CRS}.
                The
                                corrected
                                formula is indeed equivalent to the one given in \cite{CRS}
                when
                                $Z<0$, but not when $Z>0$. We believe that the formula given
                here is
                                the
                                correct one for all $Z$ because it has the expected property
                that
                                $Z\rightarrow -Z$ yields the same formula but with $Z_\infty
                                \rightarrow
                                -Z_\infty$ and
                                $\nu\rightarrow 1-\nu$.}
                                \be
                                Z = Z_\infty - {L^3\over \rho^2}\left[ {Z\over \sqrt{\rho^2 +
                Z^2}} +
                                1 -
                                2\nu\right].
                                \ee
                                Given the freedom of sign
                                for $Z_\infty$ one may again take $\nu \ge1/2$  without loss
                of
                                generality.
                                Analysis of the function $Z(\rho)$ for various choices of the
                                constants
                                $Z_\infty$ and $\nu$ yields results that are qualitatively
                similar to
                                those
                                of the D5-D3 case. The analogue of the function (\ref{dtheta})
                                turns out to be \cite{CRS}
                                \be
                                D_{\nu}(\theta) = NT_2\left[\nu - {1\over2}\left(1-\cos\theta
                                -\cos\theta\sin^2\theta\right)\right],
                                \ee
                                where $T_2$ is the M2-brane tension and (as in the D5-D3 case)
                                $\tan\theta = \rho/Z$. Although this is a quite different
                function of
                                $\theta$ from the one of (\ref{dtheta}), it has the same
                property,
                                \be
                                D_\nu(\pi) = D_{\nu-1}(0)\, ,
                                \ee
                                that is crucial to the worldvolume realization of the s-rule.

                                For $\nu=1/2$ the M5-brane interpolates between a Minkowski
                vacuum
                                embedded
                                in the M-theory vacuum and an $adS_3\times S^3$ vacuum
                embedded in
                                $adS_7\times S^4$, so this is another example of vacuum
                                interpolation `on the brane'. One would again expect an
                enhanced
                                supersymmetry
                                associated to some $adS_3$ supergroup with 16 supersymmetries.
                The
                                simplest
                                candidate supergroup is $OSp(4|2)\times OSp(4|2)$, but there
                are other
                                possibilities; we leave to the future the verification of
                enhanced
                                supersymmetry in this case and the determination of the
                precise
                                invariance supergroup.

                                \acknowledgments

                                We thank Michael Green and Alfonso Ramallo for helpful
                discussions.
                                PKT thanks ICREA for financial support during an extended
                visit to the
                                University of Barcelona, where this work was begun, and the
                members of
                                the Faculty of Physics for their hospitality. MNRW thanks the
                                University of
                                Barcelona for hospitality and financial support; he also
                thanks the
                                Gates Cambridge Trust for financial support.
                                JG is partially supported by MCYT FPA,2001-3598 and CIRIT,GC
                                2001SGR-00065.

                                
                                \end{document}